\documentclass[fleqn,usenatbib]{mnras}

\usepackage{newtxtext,newtxmath}

\usepackage[T1]{fontenc}

\DeclareRobustCommand{\VAN}[3]{#2}
\let\VANthebibliography\thebibliography
\def\thebibliography{\DeclareRobustCommand{\VAN}[3]{##3}\VANthebibliography}

\usepackage{graphicx}   
\usepackage{amsmath}    
\usepackage{amssymb}    
\usepackage{bm}

\newcommand{\ps}{\ {\rm s}^{-1}}
\newcommand{\Msun}{M_{\odot}}
\newcommand{\cm}{\ {\rm cm}}
\newcommand{\erg}{\ {\rm erg}}
\newcommand{\uG}{\ \mu{\rm G}}

\newcommand{\gray}{$\gamma$-ray}
\newcommand{\grays}{$\gamma$-rays}

\newcommand{\gamb}{\gamma_{\rm bre}}
\newcommand{\gamc}{\gamma_{\rm cut}}
\newcommand{\We}{W_{\rm e}}
\newcommand{\Wp}{W_{\rm p}}
\newcommand{\Ep}{E_{\rm p}}
\newcommand{\ee}{\eta_{\rm e}}
\newcommand{\eb}{\eta_{\rm B}}
\newcommand{\ep}{\eta_{\rm p}}

\newcommand{\hess}{{\rm H.E.S.S.}}
\newcommand{\Fermi}{{\sl Fermi}}
\newcommand{\ASg}{Tibet AS+MD}

\title[Ions in the Crab nebula]
{Modeling the broadest spectral band of the Crab nebula and constraining the ions acceleration efficiency}

\author[X. Zhang et al.]
{
Xiao Zhang,$^{1,2,}$\thanks{E-mail: xiaozhang@nju.edu.cn}
Yang Chen,$^{1,2,}$\thanks{E-mail: ygchen@nju.edu.cn}
Jing Huang$^{3}$
and Ding Chen$^{4}$ 
\\
$^{1}$School of Astronomy \& Space Science, Nanjing University, 163 Xianlin Avenue, Nanjing~210023, China \\
$^{2}$Key Laboratory of Modern Astronomy and Astrophysics, Nanjing University, Ministry of Education, China\\
$^{3}$Key Laboratory of Particle Astrophysics, Institute of High Energy Physics, Chinese Academy of Sciences, Beijing 100049, China\\
$^{4}$National Astronomical Observatories, Chinese Academy of Sciences, Beijing 100012, China \\
}

\date{Accepted XXX. Received YYY; in original form ZZZ}

\pubyear{2015}

\begin{document}
\label{firstpage}
\pagerange{\pageref{firstpage}--\pageref{lastpage}}
\maketitle

\begin{abstract}
Although it is widely accepted that the electromagnetic spectrum from radio to very-high-energy $\gamma$-rays of pulsar wind nebulae (PWNe) originates from leptons, there is still an open question that protons (or more generally, ions) may exist in pulsar wind and are further accelerated in PWN. 
The broadband spectrum of the prototype PWN Crab, extended recently by the detection of the Tibet AS$\gamma$ and HAWC experiments above 100 TeV, may be helpful in constraining the acceleration efficiency of ions.
Here, we model the broadest energy spectrum of Crab and find that the broadband spectrum can be explained by the one-zone leptonic model in which the electrons/positrons produce the emission from radio to soft $\gamma$-rays via the synchrotron process, and simultaneously generate the GeV-TeV $\gamma$-rays through inverse Compton scattering including the synchrotron self-Compton process.
In the framework of this leptonic model, the fraction of energy converted into the energetic protons is constrained to be below $0.5\ (n_{\rm t}/10\cm^{-3})^{-1}$ per cent, where $n_{\rm t}$ is the target gas density in the Crab. However, this fraction can be up to $7\ (n_{\rm t}/10\cm^{-3})^{-1}$ per cent if only the $\gamma$-rays are used.
\end{abstract}



\section{Introduction}

It is a concensus that pulsar wind nebulae (PWNe) are efficient accelerators for the extremely relativistic leptons (electrons/positrons).
A pulsar located in a PWN releases its rotational energy by driving an ultra-relativistic magnetized wind composed of electron-positron pairs.
The pulsar wind interacts with the ambient medium and slows down at the so-called termination shock where the acceleration of the pulsar wind leptons occurs \citep{Gaensler2006,kargaltsev2015}.
It is generally believed that the accelerated leptons generate the electromagnetic emission from radio to very-high-energy \grays\ via synchrotron and inverse Compton (IC, probably including synchrotron self-Compton process) mechanism, namely the leptonic model \citep[e.g.,][]{Kennel1984a,Venter2007,Zhang2008.PWN,Tanaka2010,Bucciantini2011,Martin2012,Torres2014}.

Although the leptonic model can well explain the observational features, an open question still remains whether, or how large a fraction of, protons (or more generally, ions) may exist in pulsar wind and are further accelerated in PWN.
Theoretically, it has been suggested that protons can also be extracted from the surface of a neutron star and injected into the pulsar wind \citep{Cheng1990, Arons1994}.
Protons carrying a substantial amount of the wind energy were also predicted by the resonant cyclotron absorption model \citep{Hoshino1992,Gallant1994}.
Moreover, ion acceleration in pulsars/PWNe was explored in some recent works \citep[e.g.][]{Fang2013.PSR,Chen2014.PSR,Philippov2014,Kotera2015,Lemoine2015.PWN,Guepin2020}, suggesting that pulsars/PWNs may be the sources of Cosmic Rays.
Especially, \citet{Guepin2020} concluded that the energetic protons accelerated in the pulsar magnetosphere via the magnetic reconnection can take away the spin-down energy with a fraction of 0.2 to 4.0 per cent.

These theoretical works on proton acceleration in pulsars or PWNe indicate that the \gray\ emission of a PWN may contain a contribution from hadrons.
Due to the difficulties of differentiating the leptonic and hadronic origin of the \gray\ emission, the most direct method to test this is the neutrino experiments.
With the current stacked IceCube data toward 35 \gray-bright PWNe, it can only be inferred that PWNe are not the hadronic-dominated sources \citep{Aartsen2020.PWN}.
For some individual sources, however, the hadronic component was indeed invoked to explain the \gray\ emission, e.g. Crab \citep{Atoyan1996}, Vela X \citep{Horns2006,Zhang2009.VelaX}, PWN G54.1+0.3 \citep{Li2010.G54}, and DA 495 \citep{Coerver2019}, although the purely leptonic model can also work, especially for the Crab nebula.

The Crab nebula, powered by pulsar PSR J0534+2200, is the best studied prototype PWN.
Based on the \citeauthor{Kennel1984a} (1984) 1D flow model, \citet{Atoyan1996} calculated the Crab's spectrum extending to the \gray\ band and showed that the pure IC emission underpredicts the observed flux, indicating possible additional contribution from protons.
Nontheless, in a multidimensional simulation for the flow structure in the Crab, \citet{Volpi2008} found that the leptonic model can explain the EGRET data.
With the updated GeV data obtained by \Fermi\ \citep{PWN.Crab.Fermi.2012}, the broadband spectrum of the Crab was widely modeled using the leptonic model \citep[e.g.][]{Zhang2008.PWN,Tanaka2010,Bucciantini2011,Martin2012}
without any need of a proton contribution.

Recently, $>100$ TeV photons from the Crab were detected by the \ASg\ \citep{PWN.Crab.ASg.2019} and HAWC \citep{PWN.Crab.HAWC.2019} experiments.
These new data have helped presenting the most complete spectrum across the electro-magnetic emission window, and may provide information on the proton acceleration.
In this paper, we model the broadest spectral data of the Crab nebula and explore the possibility that the very-high-energy \gray\ emission contains the contribution from the energetic protons.
The spectral evolution model is presented in section~\ref{sec:evol}.
In section~\ref{sec:had}, the hadronic component of \grays\ is constrained via the Markov Chain Monte Carlo (MCMC) method. Some discussions and conclusions are given in section~\ref{sec:dc}.


\section{The spectral evolution model}
\label{sec:evol}

\begin{figure}
\centering
\includegraphics[height=60mm]{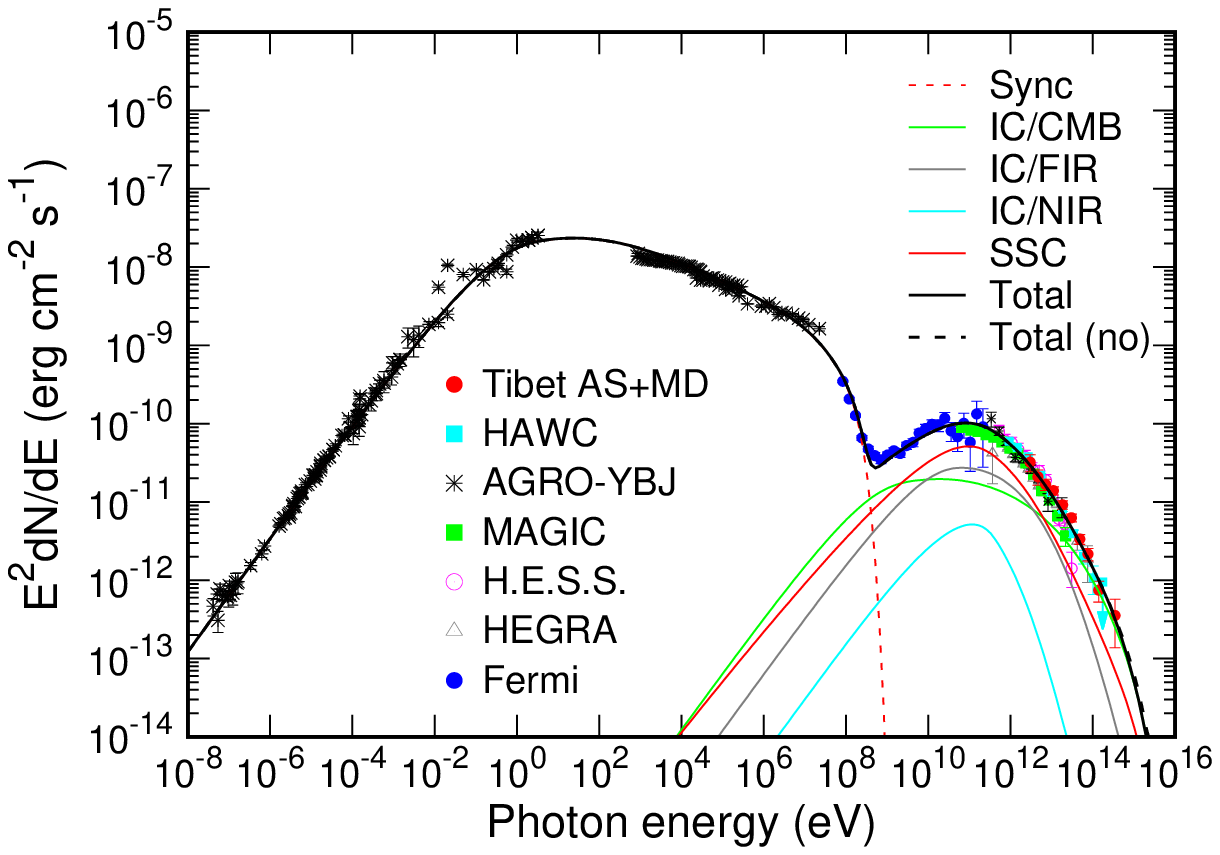}
\caption{Fit of the nonthermal spectrum of Crab for the evolutionary model.
The black solid and dashed lines show the total spectrum with and without the $\gamma-\gamma$ absorption, respectively.
The data are taken from \citet{Baldwin1971} and \citet{Macias2010} for radio band, from \citet{Ney1968}, \citet{Grasdalen1979}, \citet{Green2004}, and \citet{Temim2006} for IR, from \citet{Veron1993} for optical, from \citet{Hennessy1992} and \citet{Kuiper2001.crab} for X-rays and soft $\gamma$-rays, and from \citet[][\Fermi]{PWN.Crab.Fermi.2012}, \citet[][HEGRA]{PWN.Crab.HEGRA.2004}, \citet[][\hess]{PWN.Crab.HESS.2006}, \citet[][MAGIC]{PWN.Crab.MAGIC.2015}, \citet[][ARGO-YBJ]{PWN.Crab.ARGO.2015}, \citet[][HAWC]{PWN.Crab.HAWC.2019} and \citet[][\ASg]{PWN.Crab.ASg.2019} for GeV-TeV $\gamma$-rays.
}
\label{fig:sed_evol}
\end{figure}

The global multiband radiative properties of a PWN are generally described by the ``1-Zone model" \citep[see][for a review]{Bucciantini2014}.
The evolution of the lepton spectrum in the emission zone is given by the continuity equation in the energy space
\begin{equation}
\label{eq:dis_evol}
\frac{\partial N(\gamma,t)}{\partial t} = -\frac{\partial}{\partial\gamma}[\dot{\gamma}(\gamma,t)N(\gamma,t)] -\frac{N(\gamma,t)}{\tau(\gamma,t)} + Q(\gamma,t),
\end{equation}
where $\gamma$ is the Lorentz factor of the leptons, $N(\gamma,t)$ is the lepton distribution function, $\dot{\gamma}(\gamma,t)$ is the summation of the energy losses including the synchrotron radiation, the IC process, and the adiabatic loss \citep[see][]{Li2010.G54},
and $\tau(\gamma,t)$ represents the escape time which can be estimated via Bohm diffusion \citep[e.g.][]{Zhang2008.PWN}.
$Q(\gamma,t)$ represents the injection rate of leptons from shock acceleration process into the PWN emission zone per unit energy at a certain time and is generally assumed to be a broken power-law function
\begin{equation}
\label{eq:Qinj}
Q(\gamma,t)=Q_0(t)\left\{
\begin{array}{ll}
 (\gamma/\gamb)^{-\alpha_1}\quad {\rm for}\ \gamma<\gamb \\
 (\gamma/\gamb)^{-\alpha_2}\quad {\rm for}\ \gamma>\gamb,
\end{array}
\right.
\end{equation}
where $\gamb$ denotes the break energy, and the parameters $\alpha_1$ and $\alpha_2$ are spectral indices.
The normalization is determined from the total energy in leptons ($\We$) extracted from the spin-down energy of a pulsar ($L_{\rm sd}$) via $\We=\ee L_{\rm sd}=(1-\eb)L_{\rm sd}$, where $\ee$ and $\eb$ are the leptonic and magnetic energy fraction, respectively.
The maximum energy of the injected leptons is obtained by introducing a parameter $\varepsilon$, indicating that the Larmor radius of the lepton must be less than the termination shock radius \citep[see Eq.~(8) in][]{Martin2012}.

Solving equation~\ref{eq:dis_evol}, we can obtain the lepton distribution at a certain age.
Then we can calculate the multiwavelength nonthermal photon spectrum via the synchrotron and IC processes.
We apply this evolutionary model to Crab nebula and fit its observational multiband data including the radio \citep{Baldwin1971, Macias2010}, infra red \citep[IR,][]{Ney1968, Grasdalen1979, Green2004, Temim2006}, optical \citep{Veron1993}, X-ray \citep{Hennessy1992}, soft $\gamma$-ray, \citep{Kuiper2001.crab}, GeV $\gamma$-ray (\Fermi: \citealt{PWN.Crab.Fermi.2012}), and TeV $\gamma$-ray (HEGRA: \citealt{PWN.Crab.HEGRA.2004}; \hess: \citealt{PWN.Crab.HESS.2006}; MAGIC: \citealt{PWN.Crab.MAGIC.2015}; ARGO-YBJ: \citealt{PWN.Crab.ARGO.2015}; \ASg: \citealt{PWN.Crab.ASg.2019}) data. 
For the IC process, the seed photon fields include (i) the 2.7-K cosmic microwave background (CMB) radiation; 
(ii) the excess FIR radiation with a temperature of 70 K and an energy density of 0.5 ${\rm eV\ cm^{-3}}$ \citep{Marsden1984}; (iii) the NIR radiation with a temperature of 5000 K and an energy density of 1.0 ${\rm eV\ cm^{-3}}$ \citep{Aharonian1997}; and (iv) the synchrotron radiation photon.

In the calculation, we adopt the distance $d=2.0$~kpc, the radius $R_{\rm PWN}=2.1$~pc, the initial spin-down luminosity $L_0=3.1\times10^{39}\ \erg\ps$, and the breaking index $n=2.509$ \citep{Trimble1968, Lyne1988,Taylor1993.catalog}.
The eye-fit parameters are $\alpha_1=1.55$, $\alpha_2=2.5$, $\gamb=8.5\times10^{5}$, $\eb=0.012$, and $\varepsilon=0.3$, which are similar to previous results \citep[e.g.][]{Tanaka2010, Martin2012}.
The corresponding spectral energy distribution (SED) is plotted in Figure~\ref{fig:sed_evol}, which includes the $\gamma$-ray absorption based on the cross section of $\gamma$-$\gamma$ interaction \citep{Gould1967.pair} and the interstellar radiation field in Galaxy \citep{Shibata2011}.
As can be seen, the IC process can well explain the $\gamma$-ray data up to $\sim$500 TeV and confirm the leptonic origin for the broadband emission.
According to the fitted results, at the Crab's age of 950 yr, the average magnetic field strength in the emission zone and the maximum energy of injected leptons are about $100\uG$ and 4~PeV, respectively.
If protons can also be injected into the acceleration process, their maximum energy may be expected to reach similar order.
Moreover, protons with this energy can be well confined in the nebula at the magnetic field of $100\uG$ according to the Larmor formula, and may contribute to the $\ga 100$ TeV photons via p-p interaction.
In the next section, we study the probability that protons contribute some fraction of \gray\ emission from Crab.

\section{Hadronic component?}
\label{sec:had}
\begin{figure}
\centering
\includegraphics[height=60mm]{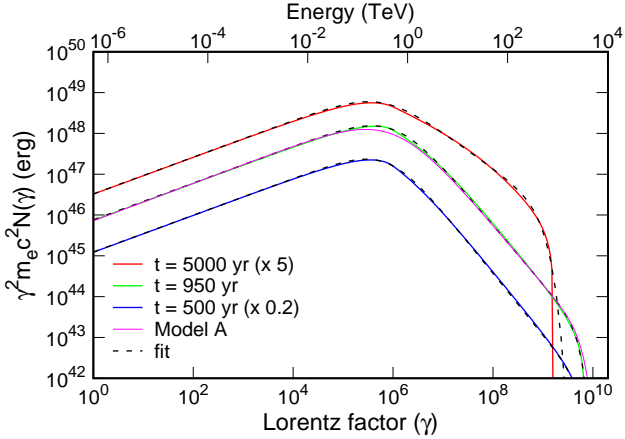}
\caption{Fit of the electron distribution in evolutionary model at three ages $t=500$, 950, and 5000 yr by using Eq.~\eqref{eq:dis_e} and 
comparison of the electron distribution at $t=950$ yr and that in Model A.
The factors enclosed in parenthesis are multiplied in plotting to distinguish the three ages.
}
\label{fig:comp}
\end{figure}

Although the broadband emissions from the Crab are explained by the leptonic process, the $\gamma$-ray emission may have hadronic component which can be constrained by the $>$100 TeV photons.
In order to quantitatively investigate this possibility, one should add the hadronic process to the above evolutionary model, and then constrain the model parameters by using a statistical method.
However, for avoiding such time-costing calculation, we can do the estimation in a simplified way.
We directly assume that the lepton distribution at the current age in the emission zone has a smooth broken power-law form with a high-energy cutoff:
\begin{equation}
\label{eq:dis_e}
\frac{dN_{\gamma}(\gamma)}{d\gamma} = A_{\rm e} \cdot \gamma^{-\alpha_1}
              \left[ 1+\left( \frac{\gamma}{\gamb} \right)^s  \right]^{\frac{\alpha_1-\alpha_2}{s}} {\rm exp}
              \left[  -\left( \frac{\gamma}{\gamc} \right)^{\beta}  \right],
\end{equation}
where $\gamc$ is the cut-off energy, $s$ the smooth parameter, and $\beta$ the cutoff shape parameter. The normalization $A_{\rm e}$ can be calculated from the total energy in leptons $\We$.
To check its accurateness, this distribution is firstly employed to fit the electrons in the evolutionary model with the above eye-fit parameters but for three different ages $t=500$, 950, and 5000 yr, which are shown in the black dotted lines in left panel of Figure~\ref{fig:comp}.
As it can be seen, the electrons in the evolutionary model can almost be described by the exponential cutoff smooth broken power law distribution.

\begin{figure*}
\centering
\includegraphics[height=60mm]{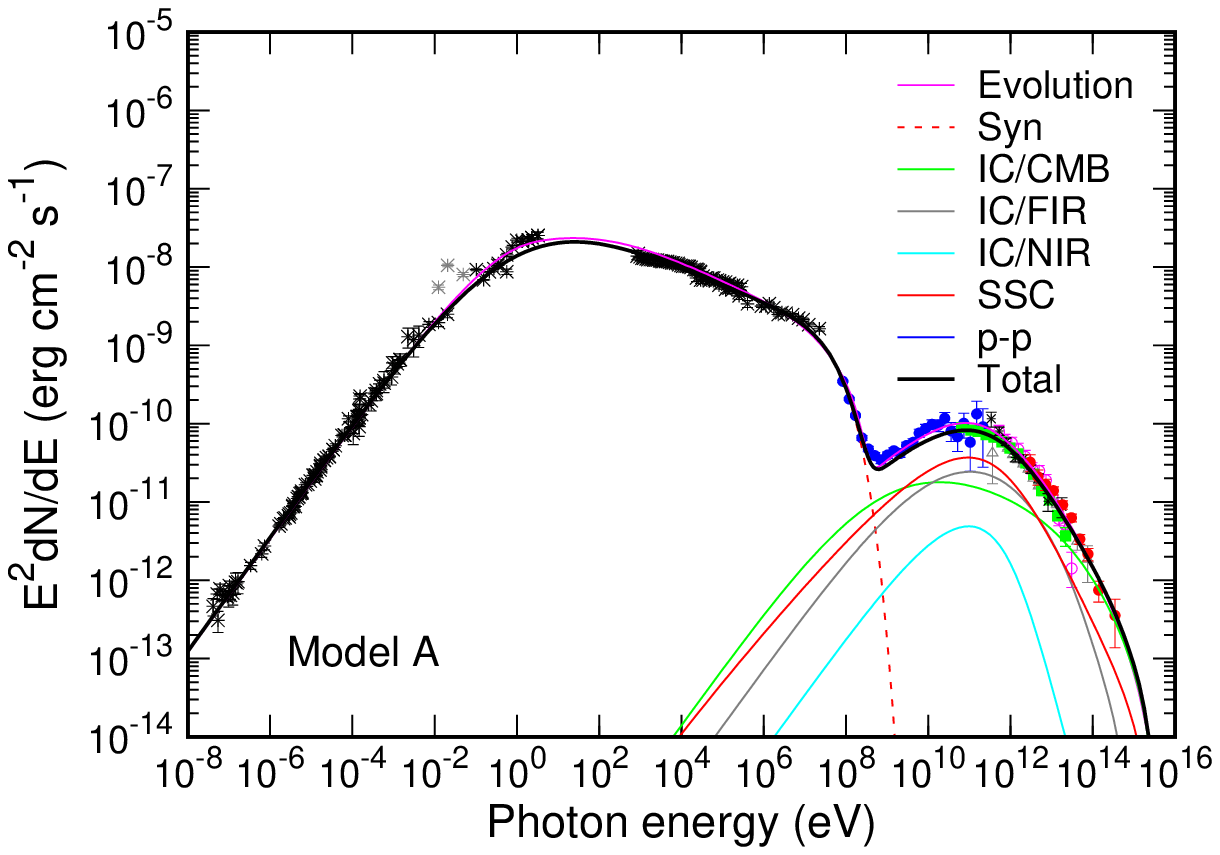}
\includegraphics[height=57mm]{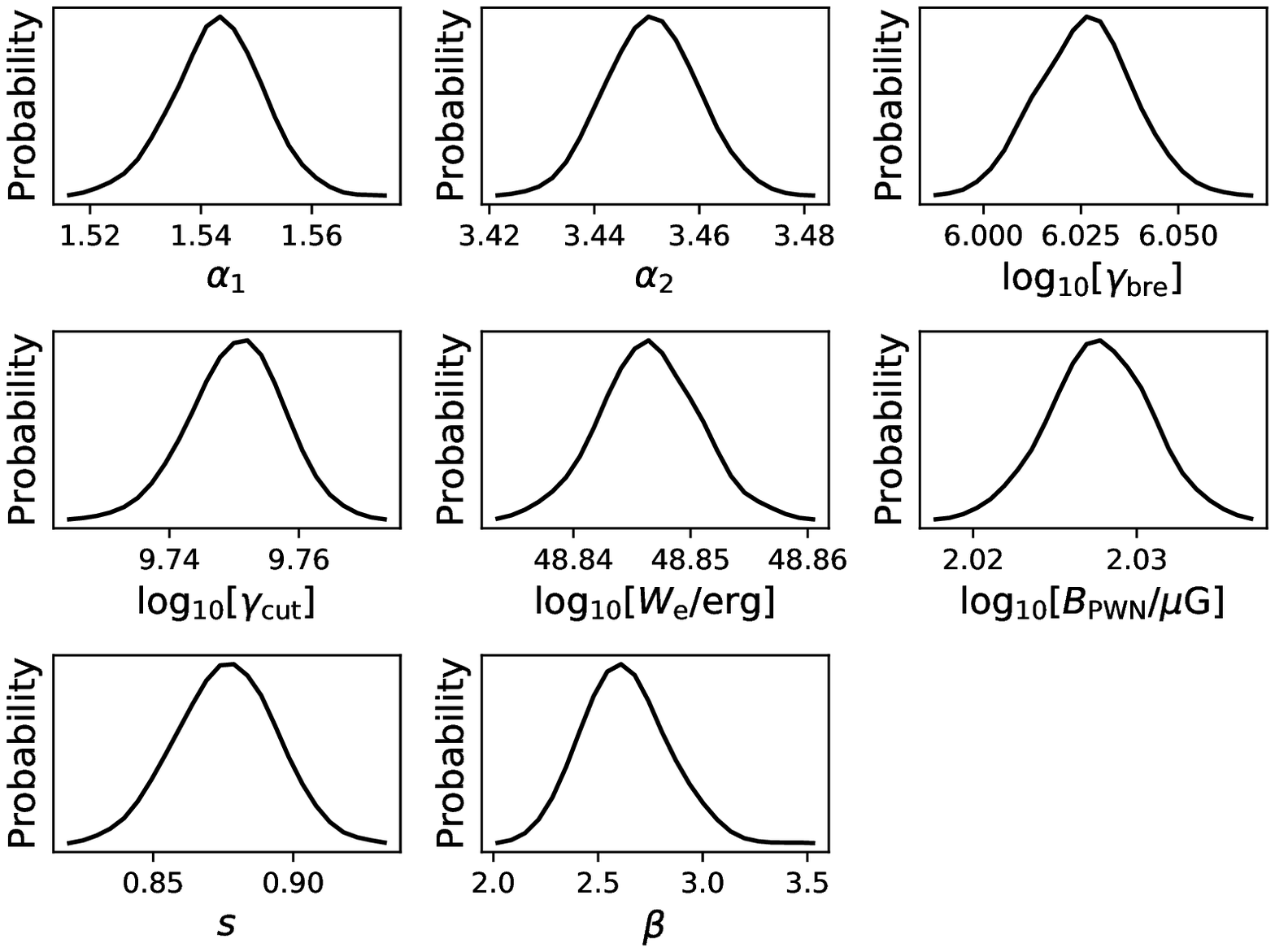}
\caption{Left: same as Figure~\ref{fig:sed_evol} but for the simplified leptonic model (Model A). The gray data in IR band are not used in the MCMC fitting. Right: corresponding 1D probability distribution of the parameters. The magenta line represents the fit in Figure~\ref{fig:sed_evol}.
}
\label{fig:sed_lep}
\end{figure*}

We now apply this simplified leptonic model to refit the broadband data of the Crab and use the MCMC approach to constrain the model parameters (Model A).
There are eight parameters in total: $\alpha_1$, $\alpha_2$, $\gamb$, $\gamc$, $B_{\rm PWN}$, $W_{\rm e}$, $s$, and $\beta$. 
The best-fit parameters are listed in Table~\ref{tab:params} and the corresponding SED is plotted in Figure~\ref{fig:sed_lep}, in which the result of the evolutionary model is also displayed in magenta.
The comparison of the electron distribution in the evolution model and that in Model A is also shown in Figure~\ref{fig:comp}.
We can see that this leptonic model is good enough to explain the broadband data, especial for the $\gtrsim$ 10 TeV data which will play an important role in constraining the hadronic component.
Thus, we will use it instead of the evolutionary model to constrain the contribution of the protons.
In addition, we can directly see that the maximum energy of the leptons at current age needs to be $\sim3$~PeV to explain the data. 
This hints that PWNe are one of potential sources of Galactic CRs.

Next, we add hadronic photons resulted from the neutral pion-decay process to the simplified leptonic model.
For the proton spectrum, we adopt the exponential cutoff power-law distribution, 
\begin{equation}
\label{eq:dis_p}
\frac{dN_{\rm p}(\Ep)}{d\Ep} = A_{\rm p} \cdot \Ep^{-\alpha_{\rm p}} {\rm exp}
              \left(  - \frac{\Ep}{E_{\rm c,p}} \right)
\end{equation}
where $\alpha_{\rm p}$ and $E_{\rm c,p}$ are the the proton index and the cutoff energy, respectively. The normalization $A_{\rm p}$ can be calculated from the total energy in protons $\Wp$.
For p-p process, we use the analytic photon emissivity developed by Kelner et al. (2006), including the enhancement factor of 1.84 due to contribution from heavy nuclei \citep{Mori2009}. 
The density of target gas interacted by relativistic protons is not well constrained.
The mean gas density in the nebula can be estimated from the total nebular mass (gas plus dust) of $7.2 \pm 0.5 \Msun$ \citep{Owen2015} and the nebular radius $R_{\rm PWN}=2.1$ pc, giving a mean density $n\sim5\ {\rm cm}^{-3}$.
Considering the fact that relativistic particles were partially captured in the dense filaments in the Crab nebula, the effective gas density for interactions of relativistic protons may be much higher than the mean density \citep{Atoyan1996}.
Here, we adopt the density of the target gas $n_{\rm t}=10\cm^{-3}$ as a fiducial value.
In addition, the nonthermal bremsstrahlung process can be neglected at this density level \citep[e.g.,][]{Atoyan1996}.

\begin{figure*}
\centering
\includegraphics[height=60mm]{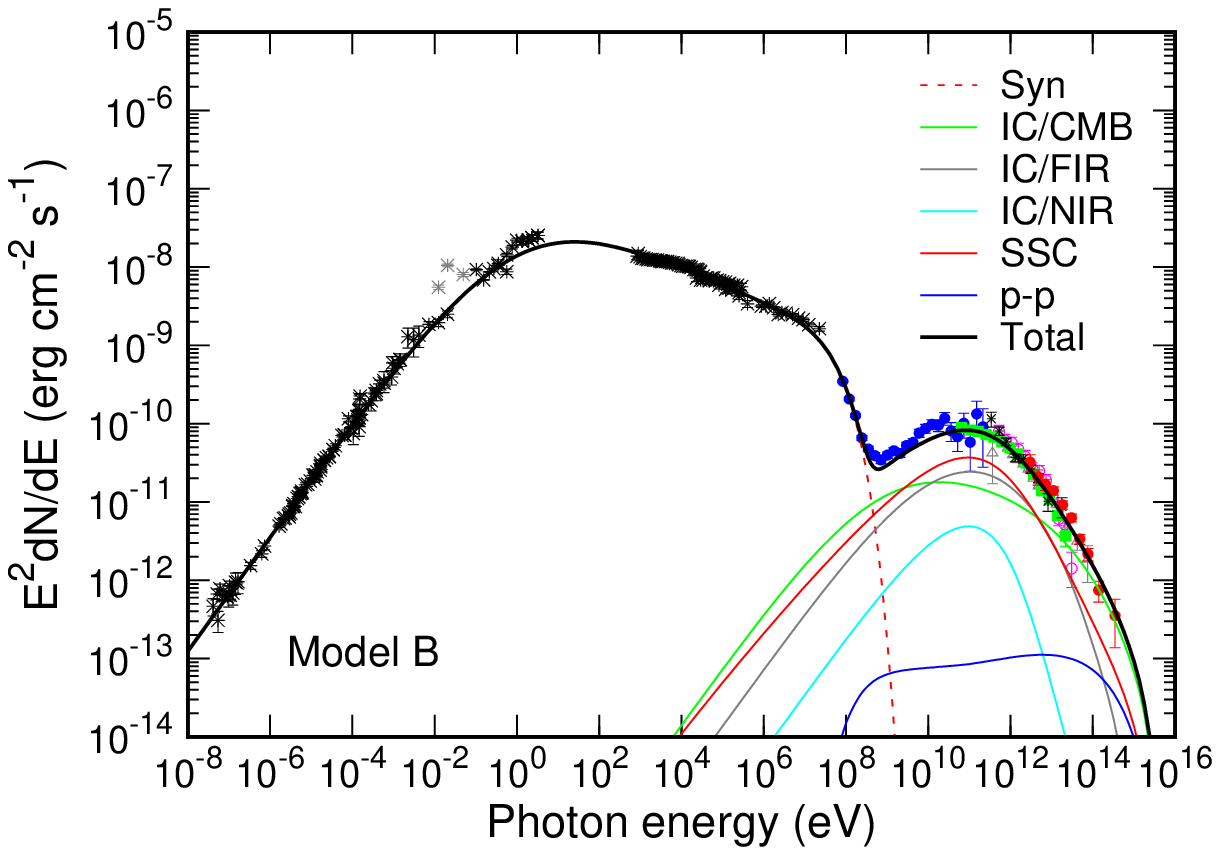}
\includegraphics[height=57mm]{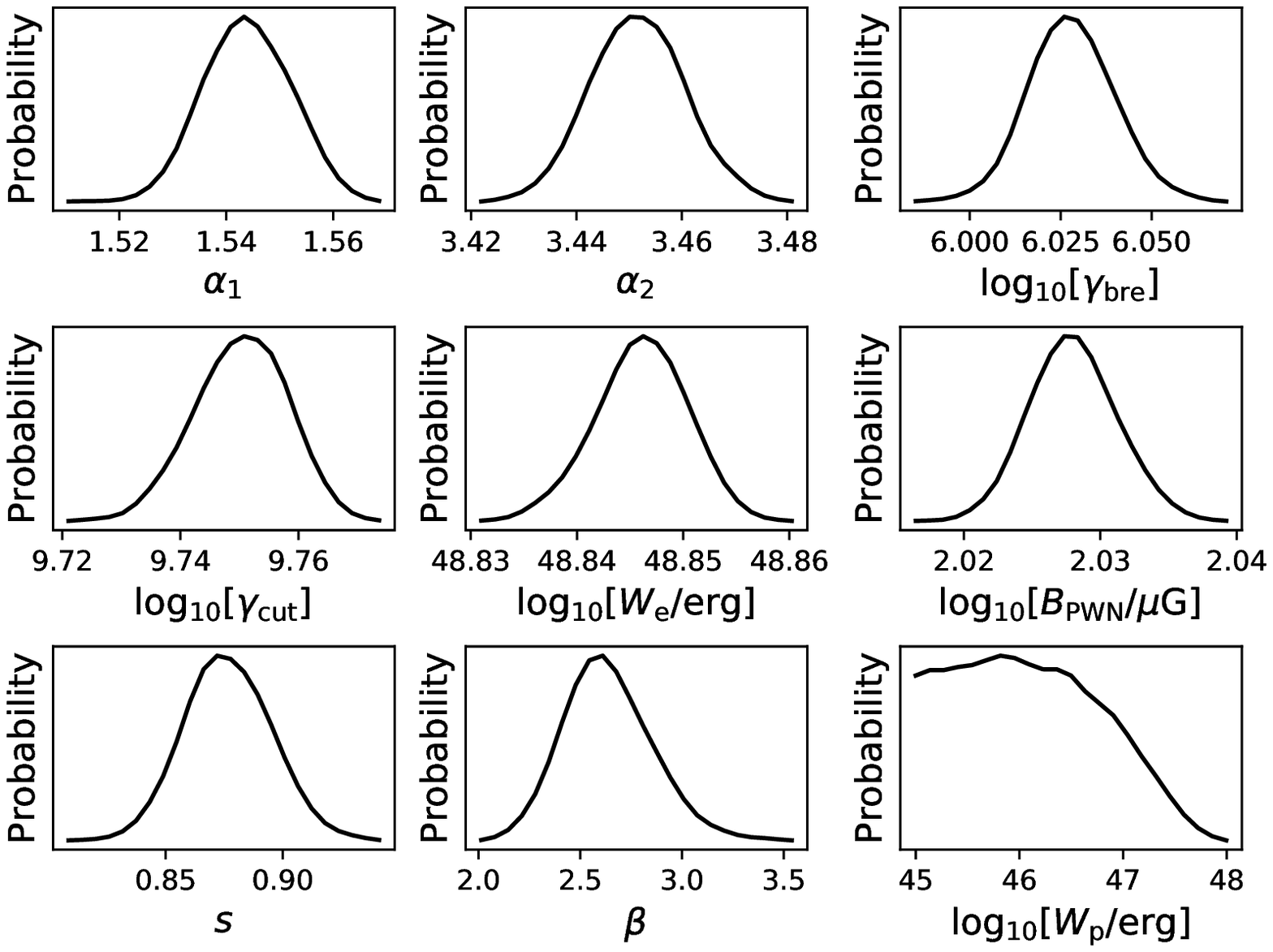}
\includegraphics[height=60mm]{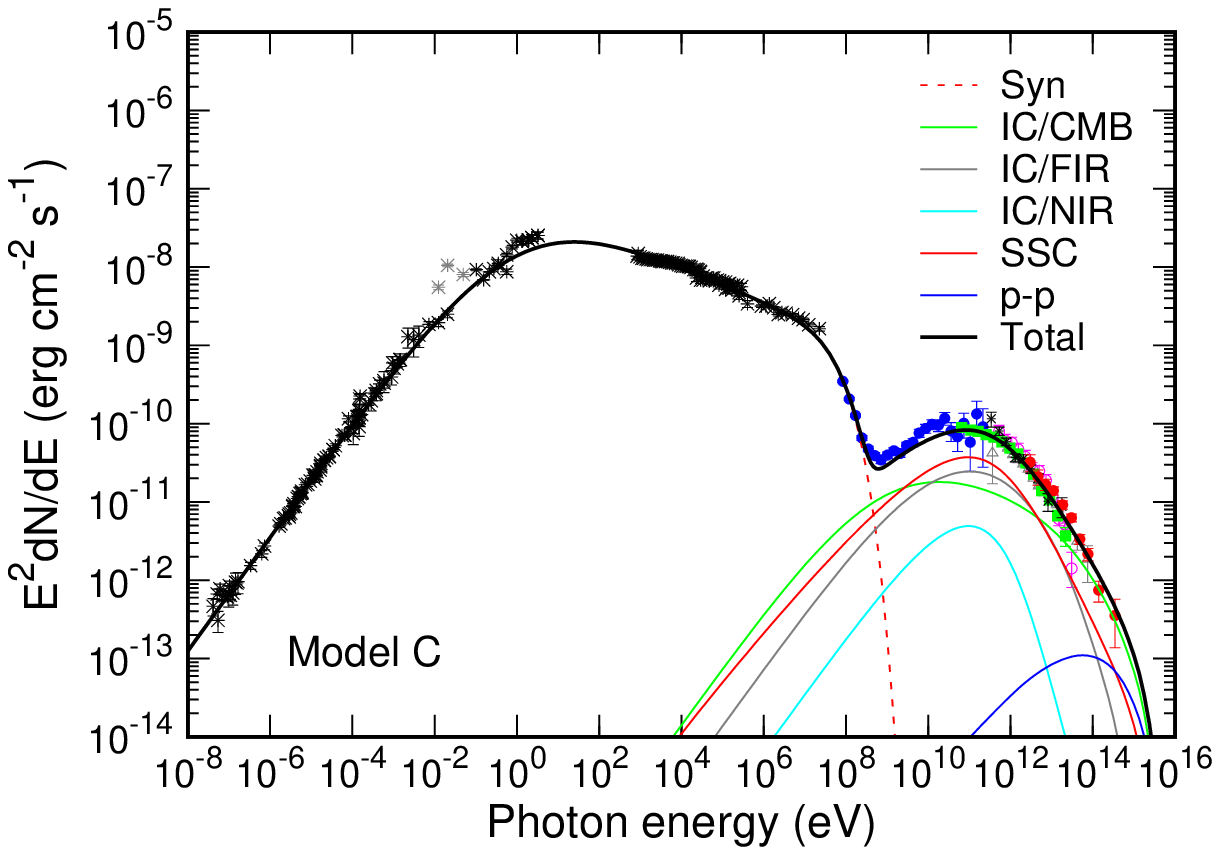}
\includegraphics[height=57mm]{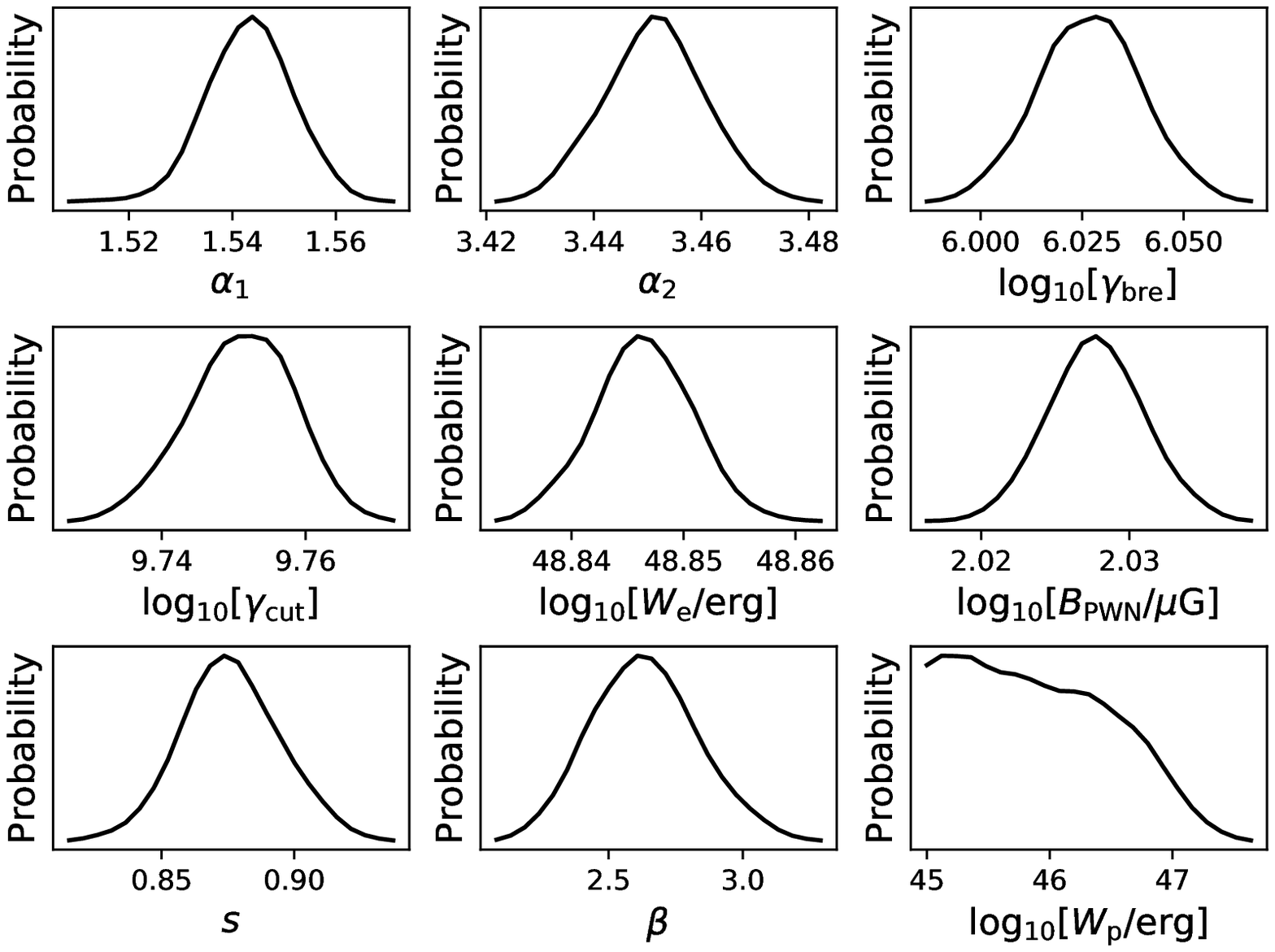}
\caption{Same as Figure~\ref{fig:sed_lep} but for Models B ($\alpha_{\rm p}=2.0$) and C ($\alpha_{\rm p}=1.55$), in which the hadronic components are considered.
}
\label{fig:sed_h1}
\end{figure*}

Due to lack of constraint on the hadronic component, we fix the proton index and the cutoff energy. 
We set $E_{\rm c,p}=3$~PeV and consider two cases for the proton index:  $\alpha_{\rm p}=2.0$ (Model B, generally assumed particle index) and $\alpha_{\rm p}=1.55$ (Model C, the same as the lepton index below the break).
For this lepto-hadronic model, there are nine parameters in total: $\alpha_1$, $\alpha_2$, $\gamb$, $\gamc$, $B_{\rm PWN}$, $\We$, $s$, $\beta$, and $\Wp$. 
The best-fit parameters are listed in Table~\ref{tab:params} and the corresponding broadband SED is plotted in Figure~\ref{fig:sed_h1}.
The fitted values of all parameters except $\Wp$ in Models B and C are almost the same as in Model A.
This means the contribution from hadronic process is negligible, which also can be seen from the result that the hadronic component (the blue line in Figure~\ref{fig:sed_h1}) is far below the observed \gray\ data.
In order to void exceeding the $> 100$~TeV data detected by \ASg, it gives a constraint on the energy in protons as $\Wp\lesssim1.9\times10^{47}$ erg and $\Wp\lesssim1.0\times10^{47}$ erg at $2\sigma$ level for $\alpha_{\rm p}=2.0$ (Model B) and $\alpha_{\rm p}=1.55$ (Model C), respectively.

\begin{table*}
 \centering
 \begin{minipage}{160mm}
  \caption{Best-fit parameters with 1$\sigma$ errors or 95 per cent upper limits. \label{tab:params}
}
  \begin{tabular}{@{}ccccccccccc@{}}
  \hline
   Model    &  $\alpha_1$  & $\alpha_2$ & log$_{10}[\gamma_{\rm bre}$] & log$_{10}[\gamma_{\rm cut}$]
   & log$_{10}[\frac{W_{\rm e}}{\rm erg}$] & log$_{10}[\frac{B_{\rm PWN}}{\rm \mu G}$] & $s$ & $\beta$ & log$_{10}[ \frac{W_{\rm p}}{\rm erg} ]$ & $\chi^2$/dof \\

 \hline
A     & $ 1.54^{+0.01 }_{-0.01 }$ & $ 3.45^{+0.01 }_{-0.01 }$ & $ 6.03^{+0.01 }_{-0.01 }$ 
      & $ 9.75^{+0.01 }_{-0.01 }$ & $48.85^{+0.00 }_{-0.00 }$ & $ 2.03^{+0.00 }_{-0.00 }$ 
      & $ 0.88^{+0.02 }_{-0.02 }$ & $ 2.63^{+0.20 }_{-0.20 }$ & --- & 523/270 \\
B     & $ 1.54^{+0.01 }_{-0.01 }$ & $ 3.45^{+0.01 }_{-0.01 }$ & $ 6.03^{+0.01 }_{-0.01 }$ 
      & $ 9.75^{+0.01 }_{-0.01 }$ & $48.85^{+0.00 }_{-0.00 }$ & $ 2.03^{+0.00 }_{-0.00 }$ 
      & $ 0.88^{+0.02 }_{-0.02 }$ & $ 2.64^{+0.20 }_{-0.20 }$ & $ \le47.27 $ & 523/269 \\
C     & $ 1.54^{+0.01 }_{-0.01 }$ & $ 3.45^{+0.01 }_{-0.01 }$ & $ 6.03^{+0.01 }_{-0.01 }$ 
      & $ 9.75^{+0.01 }_{-0.01 }$ & $48.85^{+0.00 }_{-0.00 }$ & $ 2.03^{+0.00 }_{-0.00 }$ 
      & $ 0.88^{+0.02 }_{-0.02 }$ & $ 2.63^{+0.21 }_{-0.21 }$ & $ \le47.01 $  & 523/269 \\
D     & 1.55$^a$ & $ 3.90^{+0.08 }_{-0.08 }$ & $ 6.25^{+0.07 }_{-0.08 }$ & $\ge8.76$ 
      & $49.04^{+0.03 }_{-0.04 }$ & 1.8$^a$ & 0.5$^b$ & 3.0$^b$ & $48.42^{+0.15 }_{-0.04 }$ & 79/68\\
E     & 1.55$^a$ & $ 3.83^{+0.06 }_{-0.05 }$ & $ 6.18^{+0.07 }_{-0.06 }$ & $\ge8.67$ 
      & $49.06^{+0.02 }_{-0.02 }$ & 1.8$^a$ & 0.5$^b$ & 3.0$^b$ & $48.14^{+0.19 }_{-0.06 }$ & 80/68\\
\hline
\end{tabular}
\medskip
$^a$ Eye-fit parameters.\\
$^b$ Fixed in the MCMC fit.
\end{minipage}
\end{table*}

\begin{figure*}
\centering
\includegraphics[height=60mm]{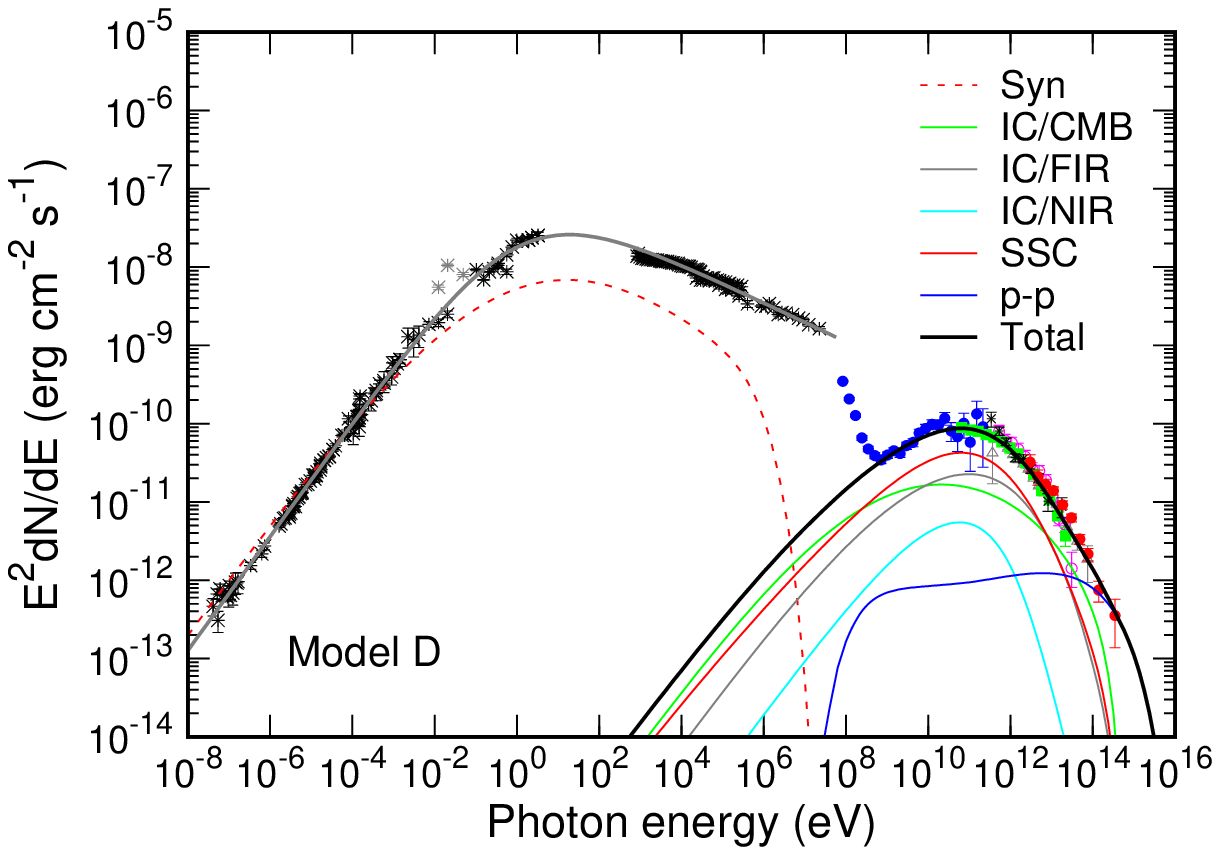}
\includegraphics[height=57mm]{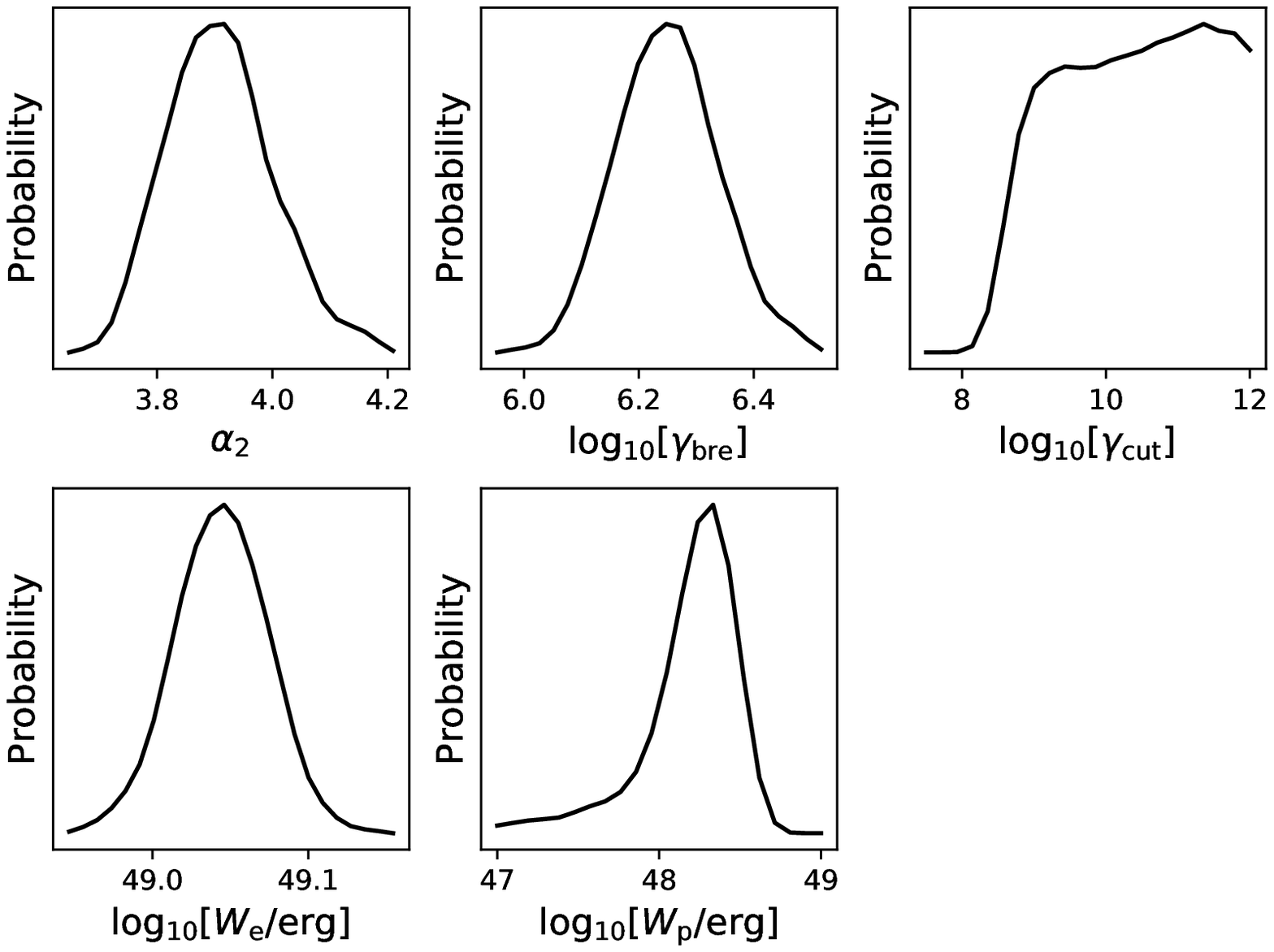}
\includegraphics[height=60mm]{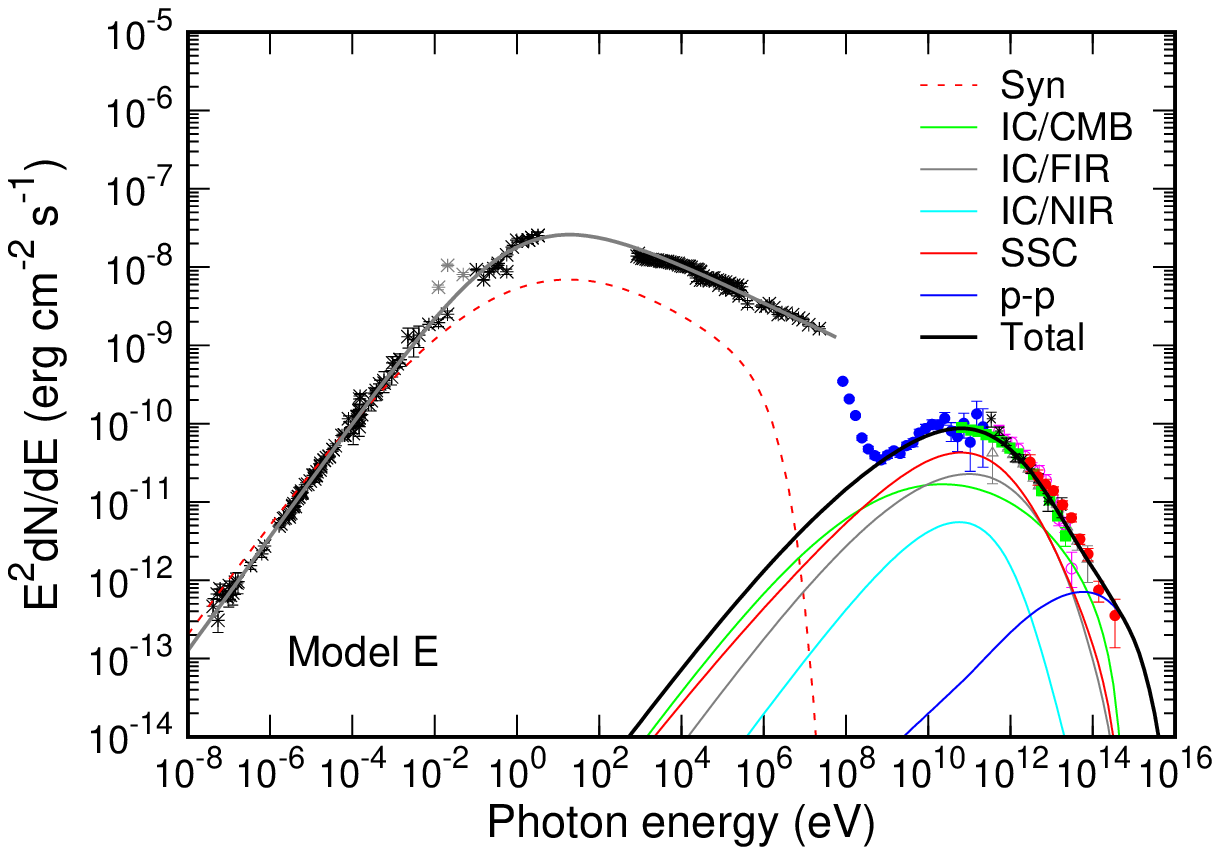}
\includegraphics[height=57mm]{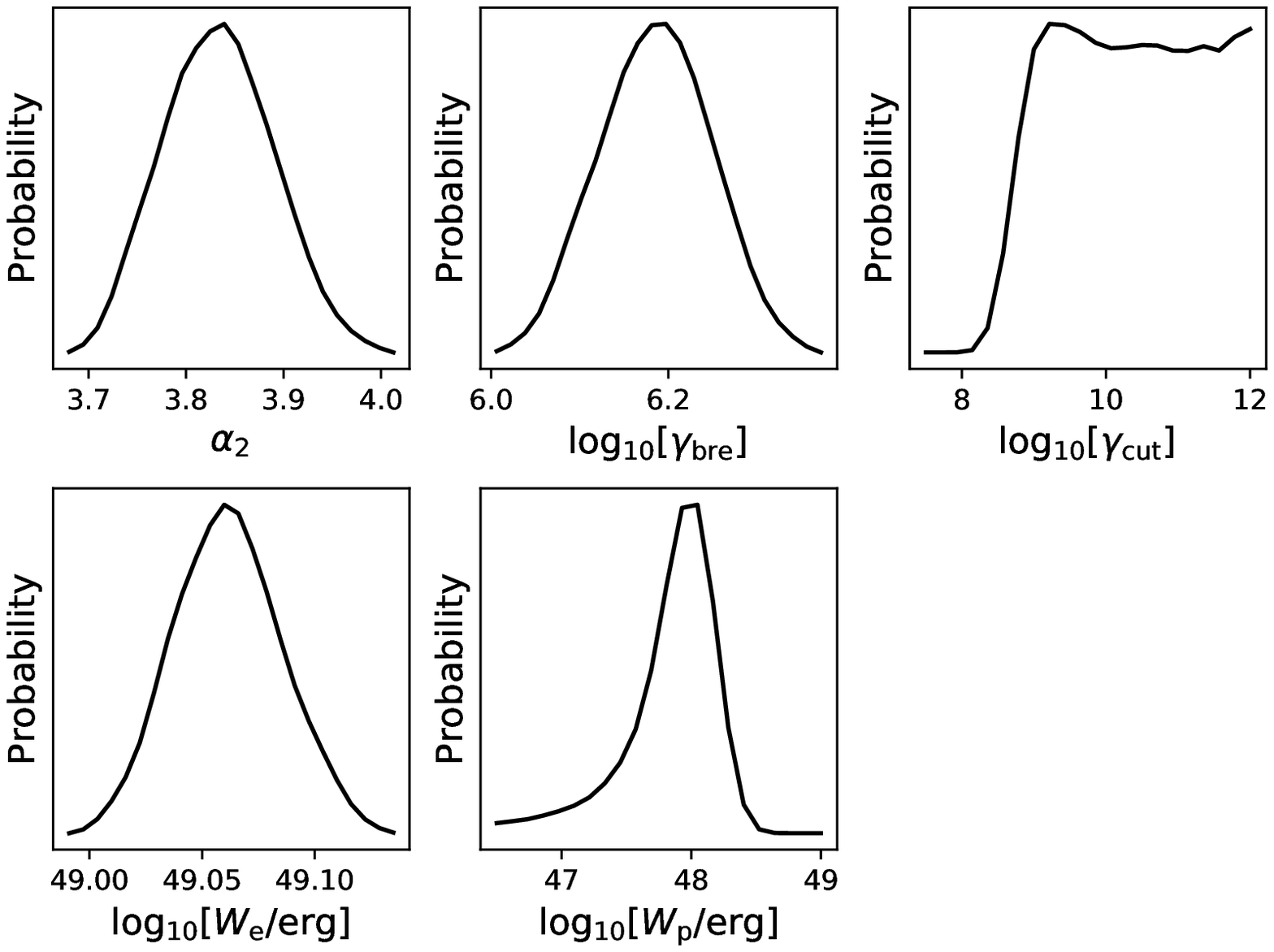}
\caption{Same as Figure~\ref{fig:sed_h1} but for Models D ($\alpha_{\rm p}=2.0$) and E ($\alpha_{\rm p}=1.55$) in which only $\gamma$-ray data with energy above 1.0 GeV are used in the MCMC fitting. The gray line represents the fit to data from radio to soft \gray, which describe the seed photons in SSC.
}
\label{fig:sed_h2}
\end{figure*}

In the calculation, we see that the residual is mainly contributed by the radio to soft \gray\ data.
To bring the \gray\ data into play,
we will only use the GeV-TeV data (i.e., above 1 GeV) to constrain the model parameters.
In doing this, we first fit the data from radio to soft \gray\ band by using a smooth broken power law function, which will describe the seed photons in the SSC process (see the gray line in left panel of Figure~\ref{fig:sed_h2}). 
Two cases $\alpha_{\rm p}=2.0$ (Model D) and $\alpha_{\rm p}=1.55$ (Model E) are also considered.

Just using \gray\ data, some of parameters do not converge very well if we do the same MCMC-fitting routine, and hence we make the following attempts.
First, the cutoff shape parameter $\beta$ is the most serious one because it is fully determined by the $\sim$100 MeV \Fermi\ data.
Considering the results in Models B and C, we round it as $\beta=3$.
In fact, due to the Klein-Nishina effect, this parameter will not obviously affect the spectral shape around 100 TeV.
Secondly, different from the narrow peak of the 1D probability in Models B and C, the smooth parameter $s$ is in a large range 0.3-0.6 and is obviously smaller that in Model B and C.
This means that the break smoothness of the leptonic spectrum is essentially determined by the IR-to-X-ray data, and the IC/\gray\ peak is broader than the synchrotron peak.
It may be the reason that the IC peak is broader than suggested by several previous models which is pointed out by MAGIC group \citep{PWN.Crab.MAGIC.2015,PWN.Crab.MAGIC.2020}.
The broader \gray\ peak not only impacts the convergence of $s$ but also $\alpha_1$, $\alpha_2$, and $\gamb$.
In order to constrain the parameters, we fix $\alpha_1=1.55$ for the index below the break and take $s=0.5$\footnote{We have also tried the case of $s=0.6$ which gives similar results.} as a example for the smooth parameter.

Then there are five parameters needed to be determined in Models D and E: $\alpha_2$, $\gamb$, $\gamc$, $\We$, and $\Wp$. 
The best-fit parameters are listed in Table~\ref{tab:params} and the corresponding SED is plotted in Figure~\ref{fig:sed_h2}.
Different from Models B and C, the hadronic component dominates the \gray\ spectrum above 100~TeV (see the blue lines in left panels of Figure~\ref{fig:sed_h2}).
This implies that the total energy in protons is constrained by the \ASg\ data as $\Wp=2.6\times10^{48}$~erg and $\Wp=1.4\times10^{48}$~erg for $\alpha_{\rm p}=2.0$ (Model D) and $\alpha_{\rm p}=1.55$ (Model E), respectively.

\section{Discussion and conclusion}
\label{sec:dc}

As shown above, the broadband spectra of the Crab nebula can be fitted with the leptonic model, in which the \gray\ emission is dominated by the IC process.
In this picture, we also constrain the contribution to \grays\ from the hadronic process.
Fitting the radio to \gray\ data via the MCMC method (namely, Models B and C), we found 
the energy in protons confined in the Crab should be less than $1.9\times10^{47}$ erg at $2\sigma$ level.
It is corresponding to the energy conversion efficiency of $\ep\sim0.5\ (n_{\rm t}/10\cm^{-3})^{-1}$ per cent, considering the total spin-down energy of  $\sim4\times10^{49}$ erg released by Crab during its whole life and the long lifetime of the energetic protons in the Crab's environment.
This fraction is far lower than the required value in the hadronic dominated models for some PWNe in which the spin-down energy are dominantly carried by ions, e.g., Vela X \citep{Horns2006} and G54.1+0.3 \citep{Li2010.G54}.
Of course, the energy carried by protons is dependent to the target gas density $n_{\rm t}$, but it can be inferred that the hadronic emissions play a minor part in explaining the \gray\ emission from the Crab in Models B and C.

\begin{figure}
\centering
\includegraphics[height=55mm]{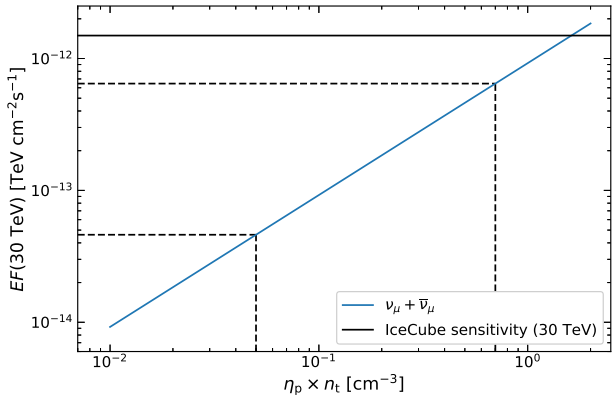}
\caption{The predicted neutrino flux from the Crab at 30 TeV as the function of $\ep\cdot n_{\rm t}$ for $\alpha_{\rm p}=2.0$ (blue). The black solid line represents the IceCube's differential sensitivity at 30 TeV adopted from \citet{Aartsen2019}. The black dashed lines are corresponding to the constraints for $\ep=0.5$ and $\ep=7$ per cent.
}
\label{fig:eta}
\end{figure}

On the other hand, in Models D and E in which only \gray\ data were used alone as an extreme case, we find that the $\sim100$~TeV \grays\ are dominated by the hadronic process.
This requires that the energy stored in protons needs to be 1.4 -- $2.6\times10^{48}$~erg, indicating that protons steal a substantial spin-down energy with a fraction of $\ep\sim7\ (n_{\rm t}/10\cm^{-3})^{-1}$ per cent.
This value is close to the upper limits of $<11\ (n_{\rm t}/10\cm^{-3})^{-1}$ per cent derived by \citet{DiPalma2017} considering the fact that IceCube has not detected any neutrinos from the Crab\footnote{It was scaled to the target density of $10 \cm^{-3}$ based on the material in PWN with the mass of $10\ M_{\odot}$ adopted in the original paper.}.
We also calculate the neutrino spectrum and find that the neutrino fluxes around 30 TeV are easy to exceed the IceCube's differential sensitivity as the increase of $\ep\cdot n_{\rm t}$.
Our predicted neutrino fluxes at 30 TeV for $\alpha_{\rm p}=2.0$ are displayed in Figure~\ref{fig:eta}, which also shows the dependence of the neutrino flux on the product of $\ep\cdot n_{\rm t}$.
It should be noted that our results are obtained on the precondition of $E_{\rm c,p}=3$~PeV which approximately equals leptons' maximum energy.
Unlike the leptons, the energy losses are negligible for protons at current age.
So their maximum energy is likely much larger than 3~PeV.
Then, in order to avoid over-estimating the flux at $\sim 100$ TeV, one should decrease the energy in protons.
Thus the conversion fraction $\ep\sim7$ per cent should be treated as an upper limits.

Recently, \citet{Xin2019.G106} performed an analysis of the Fermi data toward the TeV source VER~J2227+608 associated with a SNR-PSR complex.
Considering the fact that the GeV-TeV \gray\ spectra have a hard spectral index ($1.90\pm0.04$) and no high-energy cutoff with energy upto several dozens of TeV, they suggested that the \gray\ emission originate from the energetic protons accelerated by a PWN not a SNR.  
Based on the hadronic model, PSR J2229+6114 that is in charge of this PWN needs to convert its spin-down energy to protons with a amount of $6.0\times10^{47}(n_{\rm t}/10\cm^{-3})^{-1}$~erg during its life $\sim 10^{4}$~yr.
Considering the low spin-down luminosity of $2.2\times10^{37}\ {\rm erg\ s^{-1}}$ of PSR J2229+6114 \citep{Halpern2001.PSR}, the corresponding energy conversion efficiency of protons is $\ep\approx10$.
This value is consistent with our results given by Models D and E, or by Models B and C if the target density in Crab is down to $\sim0.5\cm^{-3}$.

In addition, for Models D and E the population of leptons determined by the \gray\ data also can produce radio to soft-\gray\ emission via synchrotron process.
In order to match the radio data, we obtain $B_{\rm PWN}\approx63\uG$ (see red dashed lines in the left panel of Figure~\ref{fig:sed_h2}).
As can be seen, this population underestimates the fluxes in the IR-to-soft-gray bands.
It implies that another population of leptons are needed to fully explain the data, which is consistent with the two-component leptonic models \citep{Zhu2015.Crab,Luo2020.Crab}.

In conclusion, the broadband spectra of Crab can be explained by the leptonic model, giving a constraint on the energy conversion fraction of protons $\ep<0.5$ per cent.
With this small fraction, however, it can not fully rule out that PWNe are PeVatrons of protons based on the current data.

\section*{Acknowledgements}
X.Z. is indebted to Siming Liu for the helpful discussions.
We thank the support of National Key R\&D Program of China under nos. 2018YFA0404204 and 2017YFA0402600 and NSFC grants under nos. U1931204, 11803011, 11851305, 11773014, 11633007, 11873065, 11673041, and 11533007.

\section*{Data availability}
No new data were generated or analysed in support of this research.


\bsp    
\label{lastpage}
\end{document}